\documentclass[aip,reprint,superscriptaddress,twocolumn,nolengthcheck]{revtex4-1}

\usepackage{graphicx}  
\usepackage{amsmath}
\usepackage{amssymb}
\usepackage{units}

\def\apj{Astrophysical Journal}
\def\apjl{Astrophysical Journal Letters}
\def\prd{Physical Review D}
\def\physrep{Physics Reports}

\begin{document}

\title{An Absence of Neutrinos Associated with Cosmic Ray Acceleration in Gamma-Ray Bursts}

\affiliation{III. Physikalisches Institut, RWTH Aachen University, D-52056 Aachen, Germany}
\affiliation{School of Chemistry \& Physics, University of Adelaide, Adelaide SA, 5005 Australia}
\affiliation{Dept.~of Physics and Astronomy, University of Alaska Anchorage, 3211 Providence Dr., Anchorage, AK 99508, USA}
\affiliation{CTSPS, Clark-Atlanta University, Atlanta, GA 30314, USA}
\affiliation{School of Physics and Center for Relativistic Astrophysics, Georgia Institute of Technology, Atlanta, GA 30332, USA}
\affiliation{Dipartimento di Fisica, Sezione INFN, I-70126, Bari, Italy}
\affiliation{Dept.~of Physics, Southern University, Baton Rouge, LA 70813, USA}
\affiliation{Dept.~of Physics, University of California, Berkeley, CA 94720, USA}
\affiliation{Lawrence Berkeley National Laboratory, Berkeley, CA 94720, USA}
\affiliation{Institut f\"ur Physik, Humboldt-Universit\"at zu Berlin, D-12489 Berlin, Germany}
\affiliation{Fakult\"at f\"ur Physik \& Astronomie, Ruhr-Universit\"at Bochum, D-44780 Bochum, Germany}
\affiliation{Physikalisches Institut, Universit\"at Bonn, Nussallee 12, D-53115 Bonn, Germany}
\affiliation{Dept.~of Physics, University of the West Indies, Cave Hill Campus, Bridgetown BB11000, Barbados}
\affiliation{Universit\'e Libre de Bruxelles, Science Faculty CP230, B-1050 Brussels, Belgium}
\affiliation{Vrije Universiteit Brussel, Dienst ELEM, B-1050 Brussels, Belgium}
\affiliation{Dept.~of Physics, Chiba University, Chiba 263-8522, Japan}
\affiliation{Dept.~of Physics and Astronomy, University of Canterbury, Private Bag 4800, Christchurch, New Zealand}
\affiliation{Dept.~of Physics, University of Maryland, College Park, MD 20742, USA}
\affiliation{Dept.~of Physics and Center for Cosmology and Astro-Particle Physics, Ohio State University, Columbus, OH 43210, USA}
\affiliation{Dept.~of Astronomy, Ohio State University, Columbus, OH 43210, USA}
\affiliation{Dept.~of Physics, TU Dortmund University, D-44221 Dortmund, Germany}
\affiliation{Dept.~of Physics, University of Alberta, Edmonton, Alberta, Canada T6G 2G7}
\affiliation{Technische Universit\"at M\"unchen, D-85748 Garching, Germany}
\affiliation{D\'epartement de physique nucl\'eaire et corpusculaire, Universit\'e de Gen\`eve, CH-1211 Gen\`eve, Switzerland}
\affiliation{Dept.~of Physics and Astronomy, University of Gent, B-9000 Gent, Belgium}
\affiliation{NASA Goddard Space Flight Center, Greenbelt, MD 20771, USA}
\affiliation{Max-Planck-Institut f\"ur Kernphysik, D-69177 Heidelberg, Germany}
\affiliation{Dept.~of Physics and Astronomy, University of California, Irvine, CA 92697, USA}
\affiliation{Laboratory for High Energy Physics, \'Ecole Polytechnique F\'ed\'erale, CH-1015 Lausanne, Switzerland}
\affiliation{Dept.~of Physics and Astronomy, University of Kansas, Lawrence, KS 66045, USA}
\affiliation{Los Alamos National Laboratory, Los Alamos, NM 87545, USA}
\affiliation{Dept.~of Astronomy, University of Wisconsin, Madison, WI 53706, USA}
\affiliation{Dept.~of Physics, University of Wisconsin, Madison, WI 53706, USA}
\affiliation{Institute of Physics, University of Mainz, Staudinger Weg 7, D-55099 Mainz, Germany}
\affiliation{Universit\'e de Mons, 7000 Mons, Belgium}
\affiliation{Bartol Research Institute and Department of Physics and Astronomy, University of Delaware, Newark, DE 19716, USA}
\affiliation{Dept.~of Physics, University of Oxford, 1 Keble Road, Oxford OX1 3NP, UK}
\affiliation{Dept.~of Physics, South Dakota School of Mines and Technology, Rapid City, SD 57701, USA}
\affiliation{Dept.~of Physics, University of Wisconsin, River Falls, WI 54022, USA}
\affiliation{Oskar Klein Centre and Dept.~of Physics, Stockholm University, SE-10691 Stockholm, Sweden}
\affiliation{Department of Physics and Astronomy, Stony Brook University, Stony Brook, NY 11794-3800, USA}
\affiliation{Dept.~of Physics and Astronomy, University of Alabama, Tuscaloosa, AL 35487, USA}
\affiliation{Dept.~of Astronomy and Astrophysics, Pennsylvania State University, University Park, PA 16802, USA}
\affiliation{Dept.~of Physics, Pennsylvania State University, University Park, PA 16802, USA}
\affiliation{Dept.~of Physics and Astronomy, Uppsala University, Box 516, S-75120 Uppsala, Sweden}
\affiliation{Dept.~of Physics, University of Wuppertal, D-42119 Wuppertal, Germany}
\affiliation{DESY, D-15735 Zeuthen, Germany}

\author{R.~Abbasi}
\affiliation{Dept.~of Physics, University of Wisconsin, Madison, WI 53706, USA}
\author{Y.~Abdou}
\affiliation{Dept.~of Physics and Astronomy, University of Gent, B-9000 Gent, Belgium}
\author{T.~Abu-Zayyad}
\affiliation{Dept.~of Physics, University of Wisconsin, River Falls, WI 54022, USA}
\author{M.~Ackermann}
\affiliation{DESY, D-15735 Zeuthen, Germany}
\author{J.~Adams}
\affiliation{Dept.~of Physics and Astronomy, University of Canterbury, Private Bag 4800, Christchurch, New Zealand}
\author{J.~A.~Aguilar}
\affiliation{D\'epartement de physique nucl\'eaire et corpusculaire, Universit\'e de Gen\`eve, CH-1211 Gen\`eve, Switzerland}
\author{M.~Ahlers}
\affiliation{Dept.~of Physics, University of Wisconsin, Madison, WI 53706, USA}
\author{D.~Altmann}
\affiliation{III. Physikalisches Institut, RWTH Aachen University, D-52056 Aachen, Germany}
\author{K.~Andeen}
\affiliation{Dept.~of Physics, University of Wisconsin, Madison, WI 53706, USA}
\author{J.~Auffenberg}
\affiliation{Dept.~of Physics, University of Wisconsin, Madison, WI 53706, USA}
% ...

\collaboration{IceCube Collaboration}
\noaffiliation

%\begin{abstract}
%Gamma-Ray Bursts (GRBs) have been proposed as a leading candidate for acceleration of ultra high-energy cosmic rays, which would be accompanied by emission of TeV neutrinos produced in $p \gamma$-interactions during acceleration in the GRB fireball. Two analyses using data from two years of the IceCube detector produced no evidence for this neutrino emission, placing strong constraints on models of neutrino and cosmic-ray production in these sources.
%\end{abstract}
%\keywords{ IceCube, Gamma-Ray Bursts, Neutrinos, Cosmic Rays}

\maketitle

%\section{Introduction}

\begin{quotation}
Very energetic astrophysical events are required to accelerate cosmic rays to above $10^{18}$ eV. Gamma-ray bursts (GRBs) have been proposed as possible candidate sources \cite{waxman95,Vietri1995,1995ApJ...449L..37M}. In the GRB fireball model, cosmic ray acceleration should be accompanied by neutrinos produced in the decay of charged pions created in interactions between the high-energy cosmic-ray protons and gamma-rays \cite{wb97}. Previous searches for such neutrinos found none, but the constraints were weak as the sensitivity was at best approximately equal to the predicted flux \cite{baikal2011,ic22_grb,Abbasi:2011qc}. Here we report an upper limit on the flux of energetic neutrinos associated with gamma-ray bursts that is at least a factor of 3.7 below the predictions \cite{wb97,rachen1998,guetta2004,grbsonprobation}. This implies that GRBs are not the only sources of cosmic rays with energies $> 10^{18}$ eV or that the efficiency of neutrino production is much lower than has been predicted.
\end{quotation}

Neutrinos from GRBs are produced in the decay of charged pions produced in interactions between high-energy protons and the intense gamma-ray background within the GRB fireball, for example in the $\Delta$-resonance process $p + \gamma \rightarrow \Delta^+ \rightarrow n + \pi^+$. When these pions decay via $\pi^+\to\mu^+ \nu_\mu$ and $\mu^+\to e^+\nu_e\bar\nu_\mu$, they produce a flux of high-energy muon and electron neutrinos, coincident with the gamma rays, and peaking at energies of several hundred TeV \cite{wb97,2008PhR...458..173B}. Such a flux should be detectable using $\unit{km}^3$-scale instruments like the IceCube neutrino telescope \cite{IceCubeNIM,muon_reconstruction} (Suppl. Fig. \ref{fig:icecube}). Due to maximal mixing between muon and tau neutrinos, neutrinos from pion decay in and around GRBs will arrive at Earth in an equal mixture of flavors. We focus here only on muons produced in $\nu_{\mu}$ charged-current interactions. As the downgoing cosmic ray muon background presents challenges for the identification of neutrino-induced muons, we achieve our highest sensitivity for upgoing (northern hemisphere) neutrinos. However, the tight constraint of spatial and temporal coincidence with a gamma-ray burst allows some sensitivity even in the southern sky. One of the two analyses presented here therefore includes southern hemisphere gamma-ray bursts during the 59-string IceCube run.

%\section{GRB sample}

The results presented here were obtained while IceCube was under construction using the 40- and 59-string configurations of the detector, which took data from April 2008 to May 2009 and from May 2009 until May 2010, respectively. During the 59-string data taking period, 190 GRBs were observed and reported via the GRB Coordinates Network \cite{GCN}, with 105 in the northern sky. Of those GRBs, 9 were not included in our catalog due to detector downtime associated with construction and calibration. Two additional GRBs were included from test runs before the start of the official 59-string run. 117 northern-sky GRBs were included from the 40-string period \cite{Abbasi:2011qc} to compute the final combined result. GRB positions were taken from the satellite with the smallest reported error, which is typically smaller than the IceCube resolution. The GRB gamma-emission start ($\mbox{T}_{\mbox{start}}$) and stop ($\mbox{T}_{\mbox{stop}}$) times were taken by finding the earliest and latest time reported for gamma emission. 

%\section{Analysis}

As in our previous study \cite{Abbasi:2011qc}, we conducted two analyses of the IceCube data. In a model-dependent search, we examine data during the period of gamma emission reported by any satellite for neutrinos with the energy spectrum predicted from the gamma-ray spectra of individual GRBs \cite{guetta2004,ic22_grb}. The model-independent analysis searches more generically for neutrinos on wider time scales, up to the limit of sensitivity to small numbers of events at $\pm$ 1 day, or with different spectra. Both analyses follow the methods used in our previous work \cite{Abbasi:2011qc}, with the exception of slightly changed event selection and the addition of the southern hemisphere to the model-independent search. Due to the large background of down-going muons from the southern sky, the southern hemisphere analysis is sensitive mainly to higher energy events (Suppl. Fig. \ref{fig:effarea_ns}). Systematic uncertainties from detector effects have been included in the reported limits from both analyses and were estimated by varying the simulated detector response and recomputing the limit, with the dominant factor the efficiency of the detector's optical sensors.

%\subsection{Model-Dependent Analysis}

\begin{figure}
  \centering
  \includegraphics[width=\linewidth]{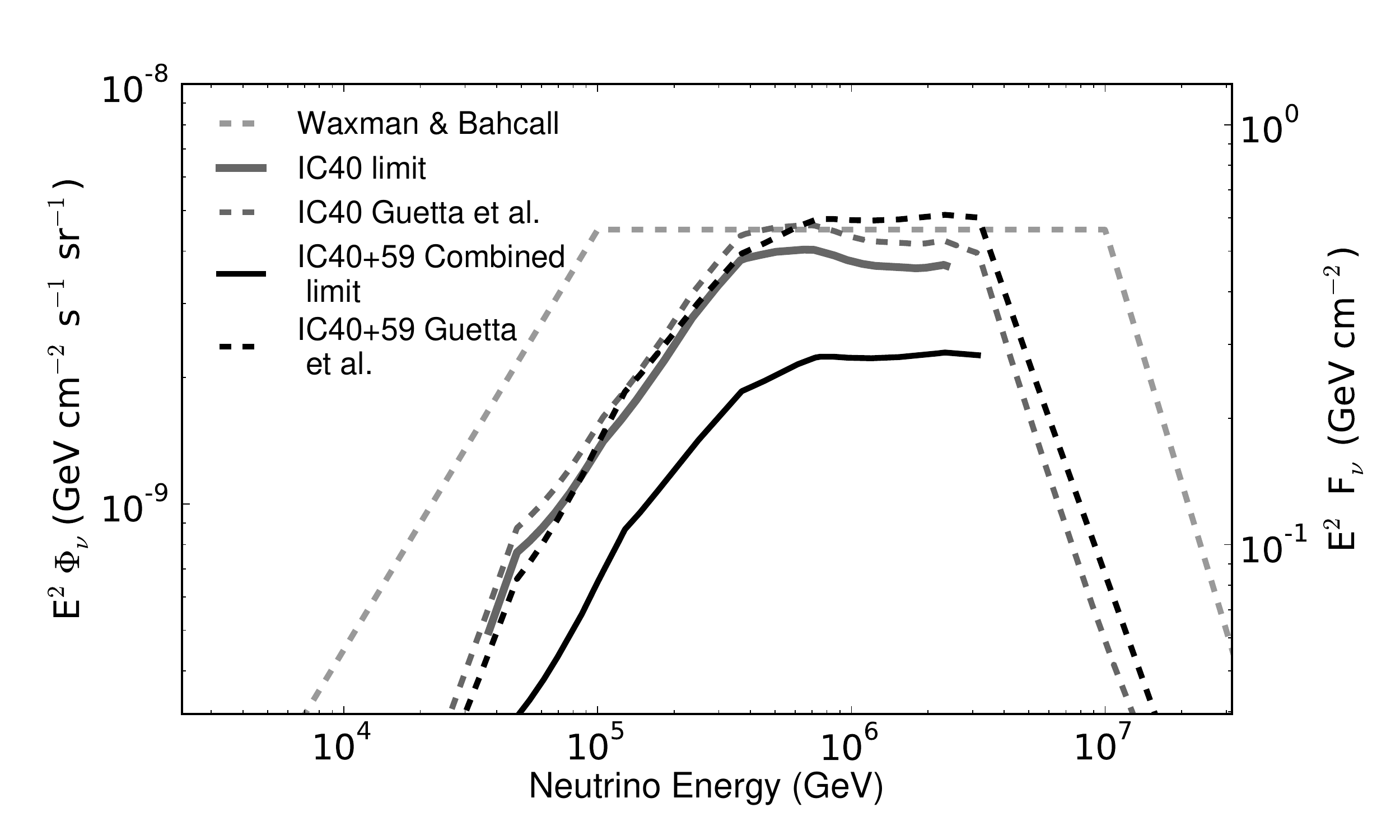}
  \caption{Comparison of results to predictions based on observed gamma-ray spectra. The summed flux predictions normalized to gamma-ray spectra \cite{guetta2004,ic22_grb,2006APh....25..118B} is shown in dashed lines; the cosmic ray normalized Waxman-Bahcall flux \cite{wb97,Waxman2003} is also shown for reference. The predicted neutrino flux, when normalized to the gamma rays \cite{guetta2004,ic22_grb}, is proportional to the ratio of energy in protons to that in electrons, which are presumed responsible for the gamma-ray emission ($\epsilon_p / \epsilon_e$, here the standard 10). The flux shown is slightly modified \cite{ic22_grb} from the original calculation \cite{guetta2004}. $\phi_{\nu}$ is the average neutrino flux at Earth, obtained by scaling the summed predictions from the bursts in our sample ($F_{\nu}$) by the global GRB rate (here 667 bursts/year \cite{Abbasi:2011qc}). The first break in the neutrino spectrum is related to the break in the photon spectrum measured by the satellites, and the threshold for photopion production, while the second break corresponds to the onset of synchrotron losses of muons and pions. Not all of the parameters used in the neutrino spectrum calculation are measurable from every burst. In such cases, benchmark values \cite{Abbasi:2011qc} were used for the unmeasured parameters. Data shown here were taken from the result of the model-dependent analysis.}
  \label{fig:combined}
\end{figure}

In the 59-string portion of the model-dependent analysis, no events were found to be both on-source and on time (within $10^\circ$ of a GRB and between $\mbox{T}_{\mbox{start}}$ and $\mbox{T}_{\mbox{stop}}$). From the individual burst spectra \cite{guetta2004,ic22_grb} with the ratio of energy in protons vs. electrons $\epsilon_p / \epsilon_e = 10$ [Ref. \citenum{ic22_grb}], 5.2 signal events were predicted from the combined 2-year dataset and a final upper limit (90\% confidence) of 0.47 times the predicted flux can be set (Fig. \ref{fig:combined}). This corresponds to a 90\% upper limit on $\epsilon_p / \epsilon_e$ of 4.7, with other parameters held fixed, and includes a $6\%$ systematic uncertainty from detector effects.

%\subsection{Model-Independent Analysis}

\begin{figure}
  \includegraphics[width=\linewidth]{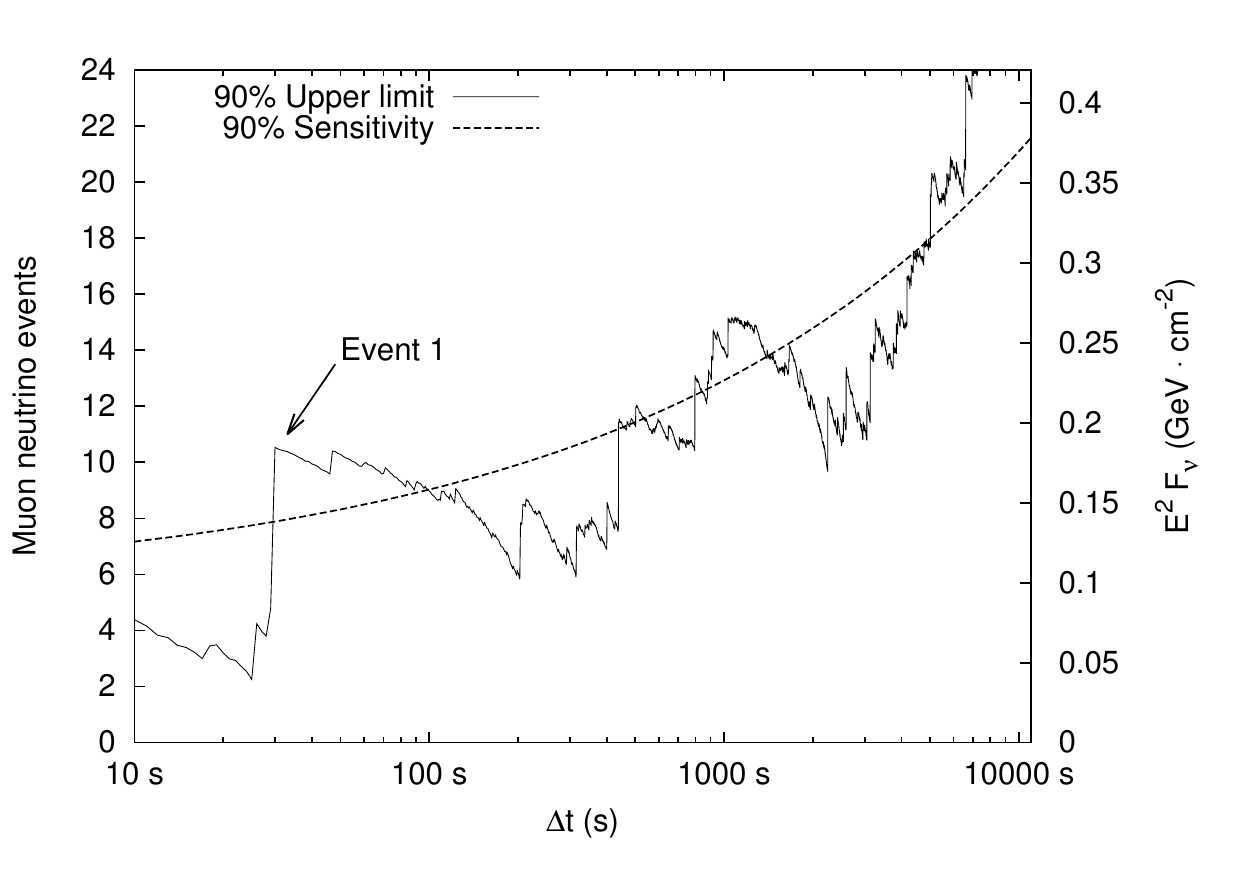}
  \caption{Upper limits on $E^{-2}$ power-law muon neutrino fluxes. Limits were calculated using the Feldman-Cousins method \cite{FeldmanCousins1998} from the results of the model-independent analysis. The left y-axis shows the total number of expected $\nu_\mu$ events while the right-hand vertical axis ($F_{\nu}$) is the same as in Fig. \ref{fig:combined}. A time window of $\Delta t$ implies observed events arriving between $t$ seconds before the burst and $t$ afterward. The variation of the upper limit with $\Delta t$ reflects statistical fluctuations in the observed background rate, as well as the presence of individual events of varying quality. The event at 30 seconds (Event 1) is consistent with background and believed to be a cosmic-ray air shower.}
  \label{fig:e2limits}
\end{figure}

In the model-independent analysis, two candidate events were observed at low significance, one 30 seconds after GRB 091026A (Event 1) and another 14 hours before GRB 091230A (most theories predict neutrinos within a few minutes of the burst). Subsequent examination showed they had both triggered several tanks in the IceTop surface air shower array, and are thus very likely muons from cosmic ray air showers.
In Fig. \ref{fig:e2limits} are shown limits from this analysis on the normalization of $E^{-2}$ muon neutrino fluxes at Earth as a function of the size of the time window $| \Delta t |$, the difference between the neutrino arrival time and the first reported satellite trigger time.
As a follow up to the model-dependent search, the limit from this analysis on the average individual burst spectra \cite{guetta2004,ic22_grb} during the time window corresponding to the median duration of the bursts in the sample (28 seconds) was 0.24 times the predicted flux, reflecting the added sensitivity of the larger burst catalog.

%\section{Discussion}

\begin{figure}
  \includegraphics[width=\linewidth]{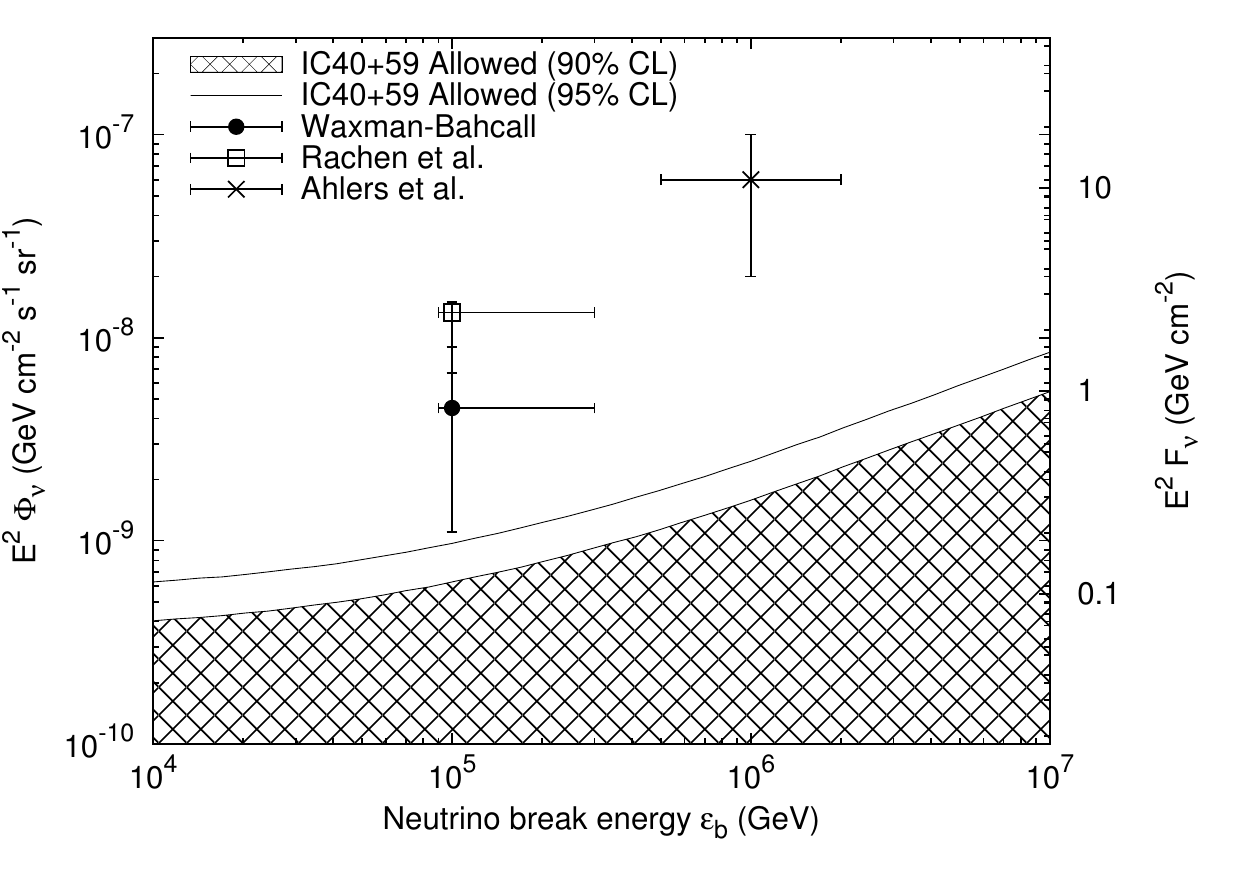}
  \caption{Compatibility of cosmic-ray flux based models with observations. Shown are the allowed values of neutrino flux vs. the neutrino break energy ($\epsilon_b$) in comparison to model predictions with estimated uncertainties. Data were taken from the model-independent analysis from the time window corresponding to the median duration of the GRBs in our catalog ($| \Delta t | = 28$ seconds).
Spectra are represented here as broken power laws ($\phi_{\nu} \cdot \{E^{-1}/\epsilon_b, E < \epsilon_b; E^{-2}, E > \epsilon_b \}$) with a break energy $\epsilon_b$ corresponding to the $\Delta$ resonance for $p \gamma$ interactions in the frame of the shock. The muon flux in IceCube is dominated by neutrinos with energies around the first break ($\epsilon_b$). As such, the upper break, due to synchrotron losses of $\pi^+$, has been neglected here, as its presence or absence does not contribute significantly to the muon flux and thus does not have a significant effect on the presented limits.
$\epsilon_b$ is related to the bulk Lorentz factor $\Gamma$ ($\epsilon_b \propto \Gamma^2$); all of the models shown assume $\Gamma \sim 300$. The value of $\Gamma$ corresponding to $10^7$ GeV is $> 1000$ for all models.
Vertical axes are related to the accelerated proton flux by the model-dependent constant of proportionality $f_{\pi}$.
For models assuming a neutron-decay origin of cosmic rays (Rachen \cite{rachen1998} and Ahlers \cite{grbsonprobation}) $f_{\pi}$ is independent of $\Gamma$; for others (Waxman-Bahcall \cite{wb97}) $f_{\pi} \propto \Gamma^{-4}$.
Error bars on model predictions are approximate and were taken either from the original papers, where included \cite{grbsonprobation}, or from the best-available source in the literature \cite{GuettaSpadaWaxman2001} otherwise. The errors are due to uncertainties in $f_{\pi}$ and in fits to the cosmic-ray spectrum.
Waxman-Bahcall \cite{wb97} and Rachen \cite{rachen1998} fluxes were calculated using a cosmic ray density of $1.5 - 3 \times 10^{44}$ erg Mpc$^{-3}$ yr$^{-1}$, with $3 \times 10^{44}$ the central value \cite{Waxman2003}.
}
  \label{fig:wballowed}
\end{figure}

Assuming that the GRBs in our catalog are a representative sample of a total of 667 per year \cite{Abbasi:2011qc}, we can scale the emission from our catalog to the emission of all GRBs. The resulting limits can then be compared to the expected neutrino rates from models that assume that GRBs are the main sources of ultra high energy cosmic rays \cite{wb97,rachen1998,grbsonprobation}, with sampling biases of the same order as model uncertainties in the flux predictions \cite{GuettaSpadaWaxman2001,aggregatesystematics}. Limits from the model-independent analysis on fluxes of this type are shown in Fig. \ref{fig:wballowed}.

These limits exclude all tested models \cite{guetta2004,wb97,rachen1998,grbsonprobation} with their standard parameters and uncertainties on those parameters (Figs. \ref{fig:combined}, \ref{fig:wballowed}). The models are different formulations of the same fireball phenomenology, producing neutrinos at proton-photon ($p \gamma$) interactions in internal shocks. The remaining parameter spaces available to each therefore have similar characteristics: either a low density of high-energy protons, below that required to explain the cosmic rays, or a low efficiency of neutrino production.

\begin{figure}
  \centering
  \includegraphics[width=\linewidth]{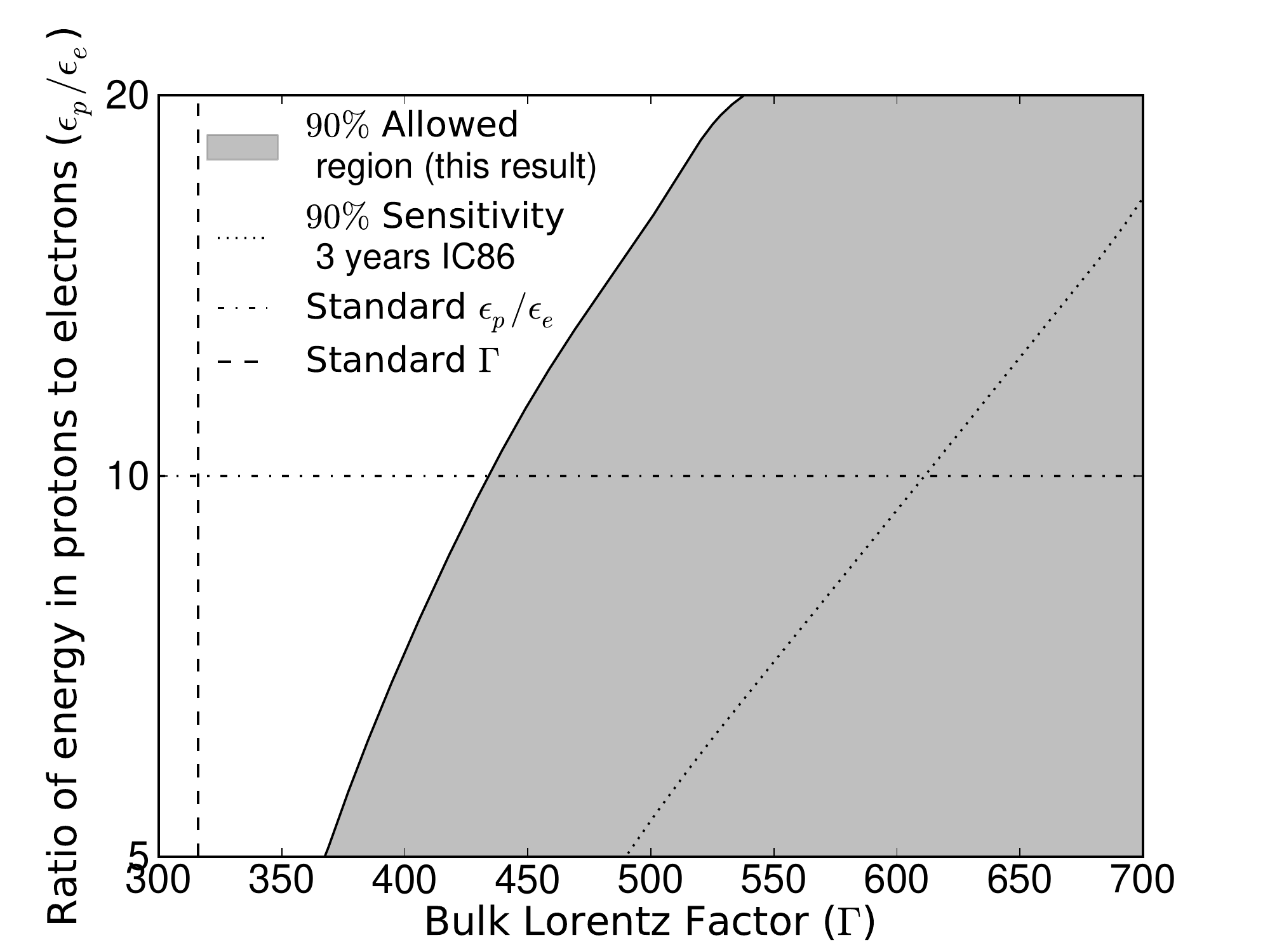}
  \caption{Constraints on fireball parameters. The shaded region, based on the result of the model-dependent analysis, shows the values of GRB energy in protons and the average fireball bulk Lorentz factor for modeled fireballs \cite{guetta2004,ic22_grb} allowed by this result at the $90\%$ confidence level. The dotted line indicates the values of the parameters to which the completed IceCube detector is expected to be sensitive after 3 years of data. The standard values considered \cite{guetta2004} are shown as dashed-dotted lines and are excluded by this analysis.  Note that the quantities shown here are model-dependent.}
  \label{fig:exclusionRegion}
\end{figure}

In the GRB fireball, protons are believed to be accelerated stochastically in collisions of internal shocks in the expanding GRB. The neutrino flux is proportional to the rate of $p \gamma$ interactions, and so to the proton content of the burst by a model-dependent factor. Assuming a model-dependent proton ejection efficiency, the proton content can in turn be related to the measured flux of high-energy cosmic rays if GRBs are the cosmic ray sources. Limits on the neutrino flux for cosmic ray normalized models are shown in Fig. \ref{fig:wballowed}; each model prediction has been normalized to a value consistent with the observed ultra high-energy cosmic ray flux. The proton density can also be expressed as a fraction of the observed burst energy, directly limiting the average proton content of the bursts in our catalog (Fig. \ref{fig:exclusionRegion}).

An alternative is to reduce the neutrino production efficiency, for example by modifying the physics included in the predictions \cite{aggregatesystematics,Hummer:2011ms} or by increasing the bulk Lorentz boost factor $\Gamma$.
Increasing $\Gamma$ increases the proton energy threshold for pion production in the observer frame, thereby reducing the neutrino flux due to the lower proton density at higher energies. Astrophysical lower limits on $\Gamma$ are established by pair production arguments \cite{guetta2004}, but the upper limit is less clear. Although it is possible that $\Gamma$ may take values of up to 1000 in some unusual bursts, the average value is likely lower (usually assumed to be around 300 \cite{guetta2004,ic22_grb}) and the non-thermal gamma-ray spectra from the bursts set a weak constraint that $\Gamma \lesssim 2000$ \cite{meszaros06}. For all considered models, with uniform fixed proton content, very high average values of $\Gamma$ are required to be compatible with our limits (Figs. \ref{fig:wballowed}, \ref{fig:exclusionRegion}).

In the case of models where cosmic rays escape from the GRB fireball as neutrons \cite{rachen1998,grbsonprobation}, the neutrons and neutrinos are created in the same $p \gamma$ interactions, directly relating the cosmic ray and neutrino fluxes and removing many uncertainties in the flux calculation. In these models, $\Gamma$ also sets the threshold energy for production of cosmic rays. The requirement that the extragalactic cosmic rays be produced in GRBs therefore does set a strong upper limit on $\Gamma$: increasing it beyond $\sim 3000$ causes the proton flux from GRBs to disagree with the measured cosmic ray flux above $4 \times 10^{18}$ eV, where extragalactic cosmic rays are believed to be dominant. Limits on $\Gamma$ in neutron-origin models from this analysis ($\gtrsim 2000$, Fig. \ref{fig:wballowed}) are very close to this point, and as a result all such models in which GRBs are responsible for the entire extragalactic cosmic-ray flux are now largely ruled out.

Although the precise constraints are model dependent, the general conclusion is the same for all the versions of fireball phenomenology we have considered here: either the proton density in gamma ray burst fireballs is substantially below the level required to explain the highest energy cosmic rays or the physics in gamma ray burst shocks is significantly different from that included in current models. In either case, our current theories of cosmic ray and neutrino production in gamma ray bursts will have to be revisited.

\bibliographystyle{naturemag}

\begin{thebibliography}{10}
\expandafter\ifx\csname url\endcsname\relax
  \def\url#1{\texttt{#1}}\fi
\expandafter\ifx\csname urlprefix\endcsname\relax\def\urlprefix{URL }\fi
\providecommand{\bibinfo}[2]{#2}
\providecommand{\eprint}[2][]{\url{#2}}

\bibitem{waxman95}
\bibinfo{author}{{Waxman}, E.}
\newblock \bibinfo{title}{Cosmological gamma-ray bursts and the highest energy
  cosmic rays}.
\newblock \emph{\bibinfo{journal}{Phys. Rev. Lett.}}
  \textbf{\bibinfo{volume}{75}}, \bibinfo{pages}{386--389}
  (\bibinfo{year}{1995}).

\bibitem{Vietri1995}
\bibinfo{author}{{Vietri}, M.}
\newblock \bibinfo{title}{{The Acceleration of Ultra--High-Energy Cosmic Rays
  in Gamma-Ray Bursts}}.
\newblock \emph{\bibinfo{journal}{\apj}} \textbf{\bibinfo{volume}{453}},
  \bibinfo{pages}{883--889} (\bibinfo{year}{1995}).
\newblock \eprint{arXiv:astro-ph/9506081}.

\bibitem{1995ApJ...449L..37M}
\bibinfo{author}{{Milgrom}, M.} \& \bibinfo{author}{{Usov}, V.}
\newblock \bibinfo{title}{{Possible Association of Ultra--High-Energy
  Cosmic-Ray Events with Strong Gamma-Ray Bursts}}.
\newblock \emph{\bibinfo{journal}{\apjl}} \textbf{\bibinfo{volume}{449}},
  \bibinfo{pages}{L37} (\bibinfo{year}{1995}).
\newblock \eprint{arXiv:astro-ph/9505009}.

\bibitem{wb97}
\bibinfo{author}{{Waxman}, E.} \& \bibinfo{author}{{Bahcall}, J.}
\newblock \bibinfo{title}{High energy neutrinos from cosmological gamma-ray
  burst fireballs}.
\newblock \emph{\bibinfo{journal}{Phys. Rev. Lett.}}
  \textbf{\bibinfo{volume}{78}}, \bibinfo{pages}{2292--2295}
  (\bibinfo{year}{1997}).

\bibitem{baikal2011}
\bibinfo{author}{{Avrorin}, A.~V.} \emph{et~al.}
\newblock \bibinfo{title}{{Search for neutrinos from gamma-ray bursts with the
  Baikal neutrino telescope NT200}}.
\newblock \emph{\bibinfo{journal}{Astronomy Letters}}
  \textbf{\bibinfo{volume}{37}}, \bibinfo{pages}{692--698}
  (\bibinfo{year}{2011}).

\bibitem{ic22_grb}
\bibinfo{author}{{Abbasi}, R.} \emph{et~al.}
\newblock \bibinfo{title}{{Search for Muon Neutrinos from Gamma-ray Bursts with
  the IceCube Neutrino Telescope}}.
\newblock \emph{\bibinfo{journal}{\apj}} \textbf{\bibinfo{volume}{710}},
  \bibinfo{pages}{346--359} (\bibinfo{year}{2010}).

\bibitem{Abbasi:2011qc}
\bibinfo{author}{Abbasi, R.} \emph{et~al.}
\newblock \bibinfo{title}{{Limits on Neutrino Emission from Gamma-Ray Bursts
  with the 40 String IceCube Detector}}.
\newblock \emph{\bibinfo{journal}{Phys. Rev. Lett.}}
  \textbf{\bibinfo{volume}{106}}, \bibinfo{pages}{141101}
  (\bibinfo{year}{2011}).
\newblock \eprint{1101.1448}.

\bibitem{rachen1998}
\bibinfo{author}{{Rachen}, J.~P.} \& \bibinfo{author}{{M{\'e}sz{\'a}ros}, P.}
\newblock \bibinfo{title}{{Cosmic rays and neutrinos from gamma-ray bursts}}.
\newblock In \bibinfo{editor}{{C.~A.~Meegan, R.~D.~Preece, \& T.~M.~Koshut}}
  (ed.) \emph{\bibinfo{booktitle}{Gamma-Ray Bursts, 4th Hunstville Symposium}},
  vol. \bibinfo{volume}{428} of \emph{\bibinfo{series}{American Institute of
  Physics Conference Series}}, \bibinfo{pages}{776--780}
  (\bibinfo{year}{1998}).
\newblock \eprint{arXiv:astro-ph/9811266}.

\bibitem{guetta2004}
\bibinfo{author}{{Guetta}, D.}, \bibinfo{author}{{Hooper}, D.},
  \bibinfo{author}{{Alvarez-Mu\~{n}iz}, J.}, \bibinfo{author}{{Halzen}, F.} \&
  \bibinfo{author}{{Reuveni}, E.}
\newblock \bibinfo{title}{{Neutrinos from individual gamma-ray bursts in the
  BATSE catalog}}.
\newblock \emph{\bibinfo{journal}{Astroparticle Physics}}
  \textbf{\bibinfo{volume}{20}}, \bibinfo{pages}{429--455}
  (\bibinfo{year}{2004}).

\bibitem{grbsonprobation}
\bibinfo{author}{{Ahlers}, M.}, \bibinfo{author}{{Gonzalez-Garcia}, M.~C.} \&
  \bibinfo{author}{{Halzen}, F.}
\newblock \bibinfo{title}{{GRBs on probation: Testing the UHE CR paradigm with
  IceCube}}.
\newblock \emph{\bibinfo{journal}{Astroparticle Physics}}
  \textbf{\bibinfo{volume}{35}}, \bibinfo{pages}{87--94}
  (\bibinfo{year}{2011}).
\newblock \eprint{1103.3421}.

\bibitem{2008PhR...458..173B}
\bibinfo{author}{{Becker}, J.~K.}
\newblock \bibinfo{title}{{High-energy neutrinos in the context of
  multimessenger astrophysics}}.
\newblock \emph{\bibinfo{journal}{\physrep}} \textbf{\bibinfo{volume}{458}},
  \bibinfo{pages}{173--246} (\bibinfo{year}{2008}).
\newblock \eprint{0710.1557}.

\bibitem{IceCubeNIM}
\bibinfo{author}{Abbasi, R.} \emph{et~al.}
\newblock \bibinfo{title}{{The IceCube data acquisition system: Signal capture,
  digitization, and timestamping}}.
\newblock \emph{\bibinfo{journal}{Nuclear Instruments and Methods in Physics
  Research A}} \textbf{\bibinfo{volume}{601}}, \bibinfo{pages}{294--316}
  (\bibinfo{year}{2009}).
\newblock \eprint{0810.4930}.

\bibitem{muon_reconstruction}
\bibinfo{author}{{Ahrens}, J.} \emph{et~al.}
\newblock \bibinfo{title}{{Muon track reconstruction and data selection
  techniques in AMANDA}}.
\newblock \emph{\bibinfo{journal}{Nucl. Instrum. and Meth. A}}
  \textbf{\bibinfo{volume}{524}}, \bibinfo{pages}{169--194}
  (\bibinfo{year}{2004}).

\bibitem{GCN}
\bibinfo{title}{{GRB Coordinates Network}}.
\newblock \bibinfo{note}{\url{http://gcn.gsfc.nasa.gov}}.

\bibitem{2006APh....25..118B}
\bibinfo{author}{{Becker}, J.~K.}, \bibinfo{author}{{Stamatikos}, M.},
  \bibinfo{author}{{Halzen}, F.} \& \bibinfo{author}{{Rhode}, W.}
\newblock \bibinfo{title}{{Coincident GRB neutrino flux predictions:
  Implications for experimental UHE neutrino physics}}.
\newblock \emph{\bibinfo{journal}{Astroparticle Physics}}
  \textbf{\bibinfo{volume}{25}}, \bibinfo{pages}{118--128}
  (\bibinfo{year}{2006}).
\newblock \eprint{arXiv:astro-ph/0511785}.

\bibitem{Waxman2003}
\bibinfo{author}{{Waxman}, E.}
\newblock \bibinfo{title}{{Astrophysical sources of high energy neutrinos}}.
\newblock \emph{\bibinfo{journal}{Nucl. Phys. B Proc. Suppl.}}
  \textbf{\bibinfo{volume}{118}}, \bibinfo{pages}{353--362}
  (\bibinfo{year}{2003}).

\bibitem{FeldmanCousins1998}
\bibinfo{author}{{Feldman}, G.~J.} \& \bibinfo{author}{{Cousins}, R.~D.}
\newblock \bibinfo{title}{{Unified approach to the classical statistical
  analysis of small signals}}.
\newblock \emph{\bibinfo{journal}{\prd}} \textbf{\bibinfo{volume}{57}},
  \bibinfo{pages}{3873--3889} (\bibinfo{year}{1998}).

\bibitem{GuettaSpadaWaxman2001}
\bibinfo{author}{{Guetta}, D.}, \bibinfo{author}{{Spada}, M.} \&
  \bibinfo{author}{{Waxman}, E.}
\newblock \bibinfo{title}{{On the Neutrino Flux from Gamma-Ray Bursts}}.
\newblock \emph{\bibinfo{journal}{\apj}} \textbf{\bibinfo{volume}{559}},
  \bibinfo{pages}{101--109} (\bibinfo{year}{2001}).

\bibitem{aggregatesystematics}
\bibinfo{author}{{Baerwald}, P.}, \bibinfo{author}{{H{\"u}mmer}, S.} \&
  \bibinfo{author}{{Winter}, W.}
\newblock \bibinfo{title}{{Systematics in Aggregated Neutrino Fluxes and Flavor
  Ratios from Gamma-Ray Bursts}}.
\newblock \emph{\bibinfo{journal}{Astroparticle Physics}}
  \textbf{\bibinfo{volume}{35}}, \bibinfo{pages}{508 -- 529}
  (\bibinfo{year}{2012}).
\newblock \eprint{1107.5583}.

\bibitem{Hummer:2011ms}
\bibinfo{author}{{H{\"u}mmer}, S.}, \bibinfo{author}{Baerwald, P.} \&
  \bibinfo{author}{Winter, W.}
\newblock \bibinfo{title}{{Neutrino Emission from Gamma-Ray Burst Fireballs,
  Revised}}  (\bibinfo{year}{2011}).
\newblock \eprint{1112.1076}.

\bibitem{meszaros06}
\bibinfo{author}{{M{\'e}sz{\'a}ros}, P.}
\newblock \bibinfo{title}{{Gamma-ray bursts}}.
\newblock \emph{\bibinfo{journal}{Reports on Progress in Physics}}
  \textbf{\bibinfo{volume}{69}}, \bibinfo{pages}{2259--2321}
  (\bibinfo{year}{2006}).
\newblock \eprint{arXiv:astro-ph/0605208}.

\end{thebibliography}

\section*{Supplementary Information}
Supplementary Information is linked to the online version of the paper at \url{www.nature.com/nature}.

\begin{acknowledgments}
We acknowledge support from the following agencies:
US NSF - Office of Polar Programs,
US NSF - Physics Division,
U. of Wisconsin Alumni Research Foundation,
the GLOW and OSG grids;
US DOE, NERSCC, the LONI grid;
NSERC, Canada;
Swedish Research Council,
Swedish Polar Research Secretariat,
SNIC,
K. and A. Wallenberg Foundation, Sweden;
German Ministry for Education and Research,
Deutsche Forschungsgemeinschaft;
Research Department of Plasmas with Complex Interactions (Bochum), Germany;
FSR,
FWO Odysseus,
IWT,
BELSPO, Belgium;
University of Oxford, United Kingdom;
Marsden Fund, New Zealand;
Australian Research Council;
JSPS, Japan;
SNSF, Switzerland;
J.~P.~R was supported by the Capes Foundation, Brazil,
N.~W. by the NSF GRFP.
Thanks to S.~H\"ummer, E.~Waxman and W.~Winter for discussions.
\end{acknowledgments}

\section*{Author Contributions}
The IceCube observatory has been designed and constructed by the IceCube Collaboration and the IceCube Project. It is operated by the IceCube Collaboration, who set science goals. Data processing and calibration, Monte Carlo simulations of the detector and of theoretical models, and data analyses are performed by a large number of IceCube members who also discuss and approve the scientific results. This manuscript was written by P.R. and N.W. and subjected to an internal collaboration-wide review process. All authors approved the final version of the manuscript.

\section*{Author Information}
Reprints and permissions information is available at \url{www.nature.com/reprints}. The authors declare no competing financial interests. Correspondence and requests for materials should be addressed to N.W. (\url{nwhitehorn@icecube.wisc.edu}) or to P.R. (\url{redlpete@icecube.umd.edu}).

\clearpage
\newpage
\appendix*

%\section*{Supplementary Information}

\ifx \standalonesupplemental\undefined
\setcounter{page}{1}
\setcounter{figure}{0}
\setcounter{table}{0}
\renewcommand{\thepage}{Supplementary Information -- S\arabic{page}}
\else
\renewcommand{\thepage}{}
\fi
\renewcommand{\figurename}{SUPPL. FIG.}

\begin{figure}
\includegraphics[width=\linewidth]{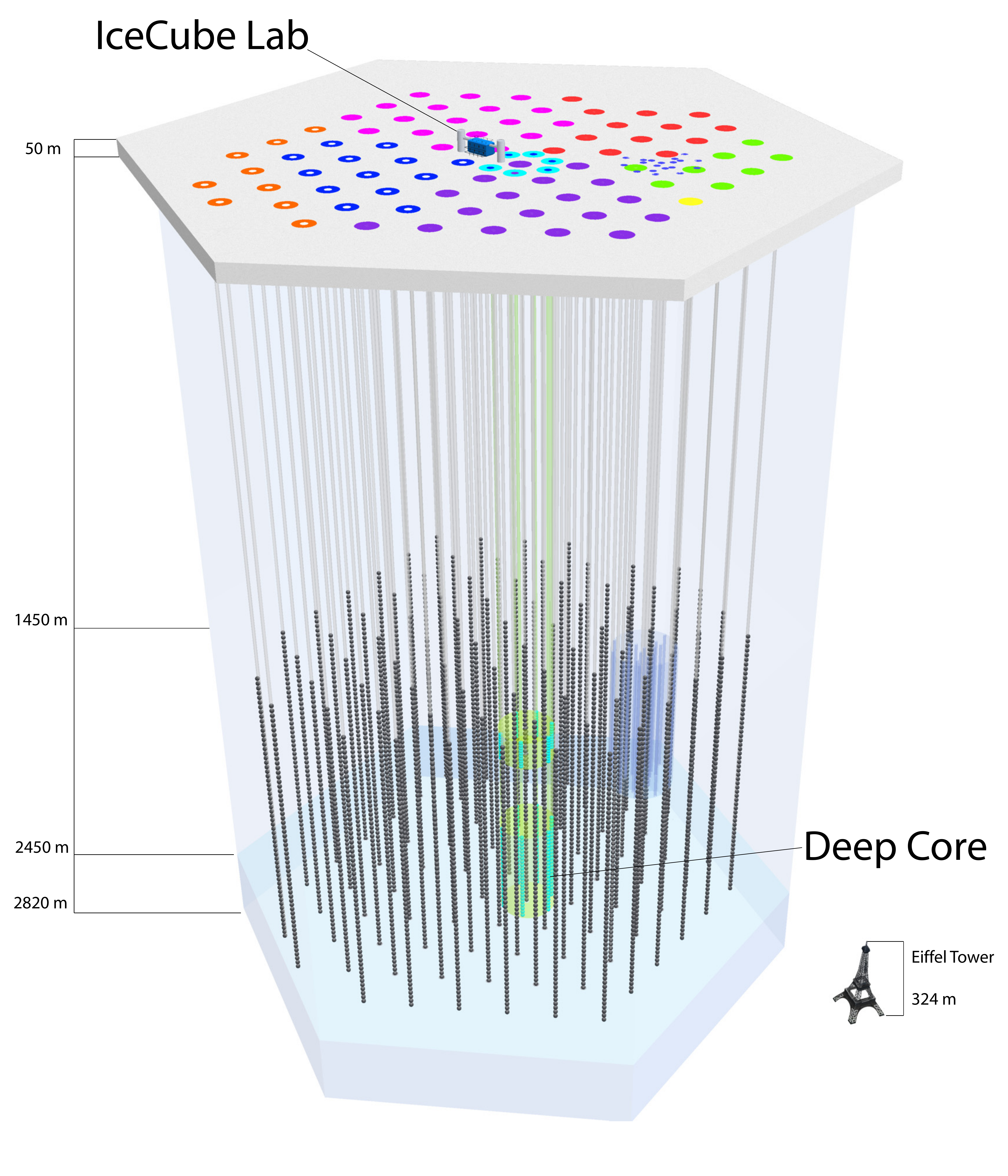}
\caption{The IceCube neutrino observatory. The IceCube detector instruments a volume of $1\,\unit{km}^3$ of glacial ice at the South Pole, sensitive to neutrinos of TeV and higher energy \cite{IceCubeNIM} (Suppl. Fig. \ref{fig:effarea_analyses}). Neutrinos are detected by observing Cherenkov light emitted by secondary charged particles produced in neutrino-nucleon interactions \cite{muon_reconstruction}, and their arrival direction is obtained from the timing pattern of the detected light.
The finished detector is composed of 5160 digital optical modules (DOMs), each containing a 10-inch photomultiplier, with 60 placed at depths between 1450 and 2450 m on each of 86 vertical strings. IceCube is complemented by a surface air shower array called IceTop \cite{IceCubeNIM}, with two tanks located above each of the IceCube strings. The colors at the top indicate the detector at various stages of deployment.
IceCube achieves its best angular resolution for muons produced in $\nu_{\mu}$ charged-current interactions ($0.6^\circ$ for $E_\nu \gtrsim 100 \ \unit{TeV}$). Combined with the increased detector effective volume afforded by the long distances traveled by the secondary muons, such events usually provide the highest sensitivity for searches for neutrino point sources.
}
\label{fig:icecube}
\end{figure}

\section*{GRB Catalog}

The GRB catalog used in this analysis was synthesized from GCN notices and can be obtained using the GRBweb database available at \url{http://grbweb.icecube.wisc.edu/}. IceCube-40 operated from April 5, 2008 until May 20, 2009 and IceCube-59 operated from May 21, 2009 until May 31, 2010. GRB090422 and GRB090423, though before the official 59-string start date, occurred during test runs of the 59-string detector and so are included in the 59-string catalog.

\section*{Effective Areas}

\begin{figure}
\includegraphics[width=\linewidth]{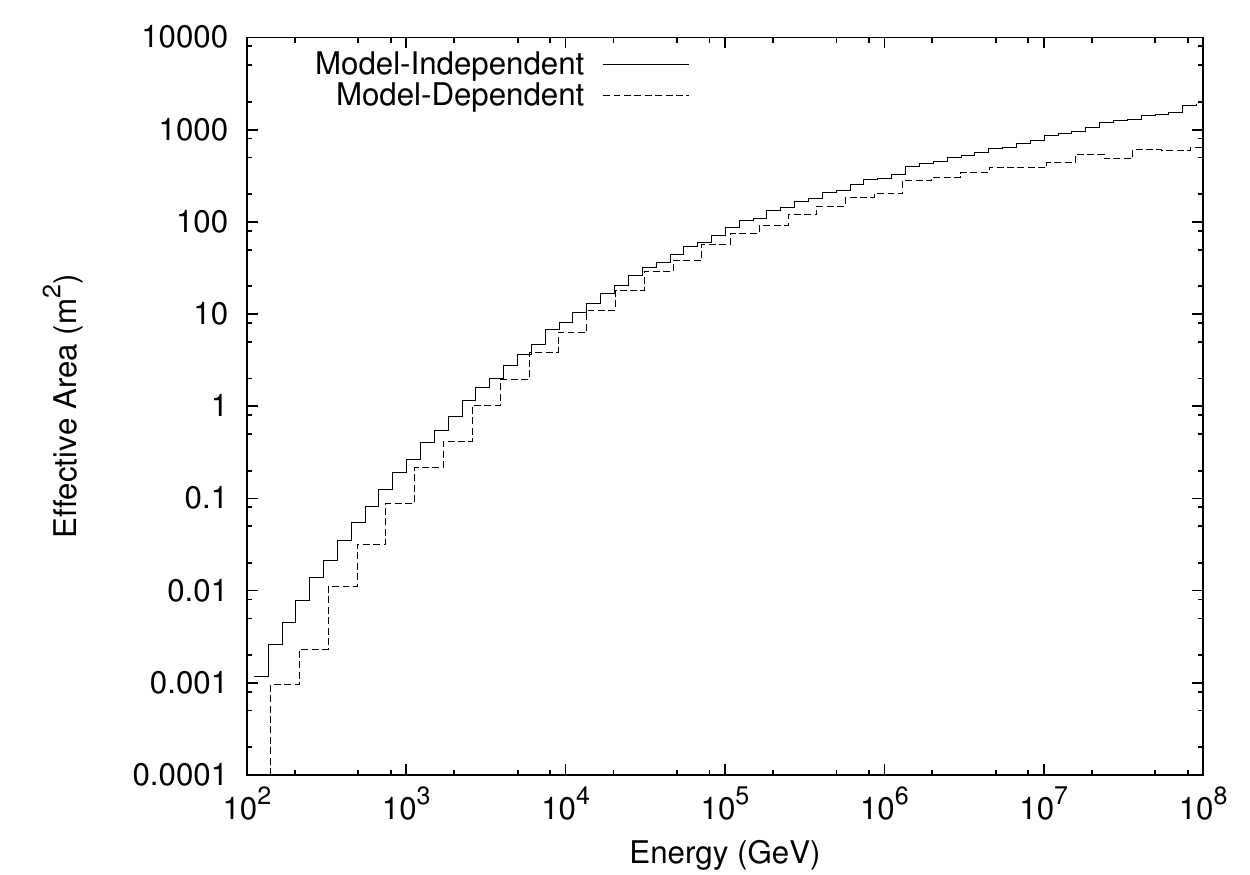}
\caption{Effective area of the IceCube neutrino telescope using the event selections of the model-dependent and model-independent analyses, averaging over the 40- and 59-string detector configurations and zenith angles according to the distribution of bursts in the catalog. The effective area of the model-independent event selection is in general somewhat larger, due to using a weight scheme instead of hard cuts -- however, the extra events so included are typically low quality and so have low weights when computing final results. The model-independent average effective area includes the southern hemisphere for the 59-string portion of the analysis (Suppl. Fig. \ref{fig:effarea_ns}).}
\label{fig:effarea_analyses}
\end{figure}

\begin{figure}
\includegraphics[width=\linewidth]{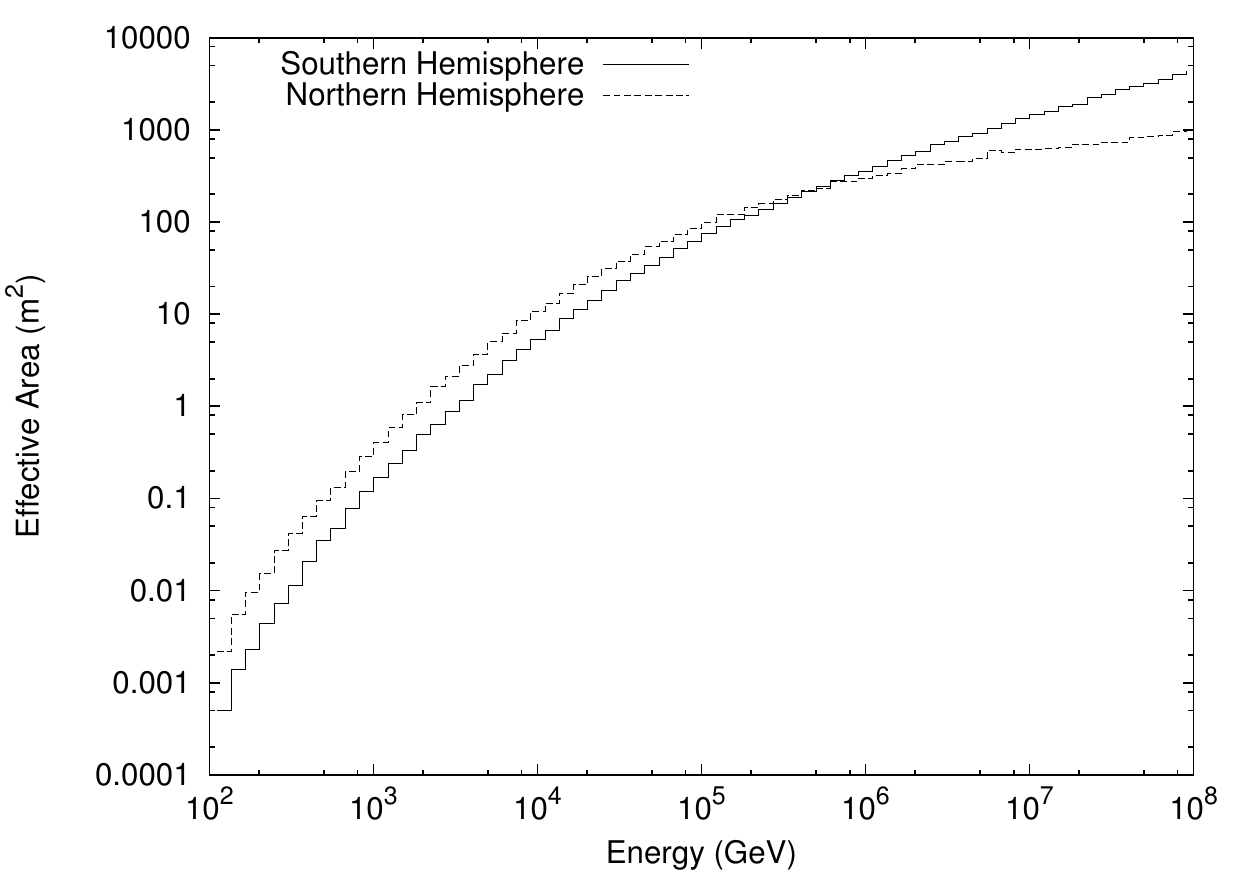}
\caption{Average effective area of the IceCube neutrino telescope in its 59-string configuration to fluxes of muon neutrinos from the northern and southern hemispheres, using the event selection from the model-independent analysis. At low energies, the ability to use the Earth to filter out neutrinos from cosmic ray air showers reduces backgrounds and improves the sensitivity of the detector to neutrinos from the northern sky. As the neutrino energy increases, so does the neutrino-nucleon cross-section, increasing the neutrino interaction probability and the chances of a detection in IceCube. At very high energies, backgrounds from cosmic-ray muons are substantially reduced and the neutrino-nucleon cross-section becomes large enough that neutrinos from the northern hemisphere will be absorbed in the Earth before reaching the detector, making the southern hemisphere the region of highest sensitivity for $E_{\nu} \gtrsim 1$ PeV. Note that the effective area is not a direct estimate of the sensitivity, due to variable backgrounds as a function of zenith angle and energy. In general, backgrounds are highest in the southern hemisphere and at low energies.}
\label{fig:effarea_ns}
\end{figure}

The detector effective areas (Suppl. Figs. \ref{fig:effarea_analyses}, \ref{fig:effarea_ns}) can be used to estimate the detector response for an arbitrary neutrino flux. Convolution of a flux with the effective area will give the expected event rate in IceCube. Presented effective areas are the average of the effective areas for muon neutrinos and muon antineutrinos and correspond to the expectation value of the detector effective area under variations to account for systematic uncertainties in the detector simulation. The increase in effective area between the 40- and 59-string detector configurations is due to the 50\% increase in geometrical area of the detector, a more favorable detector geometry, and improvements in the event selection and reconstruction techniques (Suppl. Fig. \ref{fig:effarea_ic40ic59}). Data files containing all the effective areas plotted here are included in the supplementary information (Suppl. Tables 1-6).

\begin{figure}
\includegraphics[width=\linewidth]{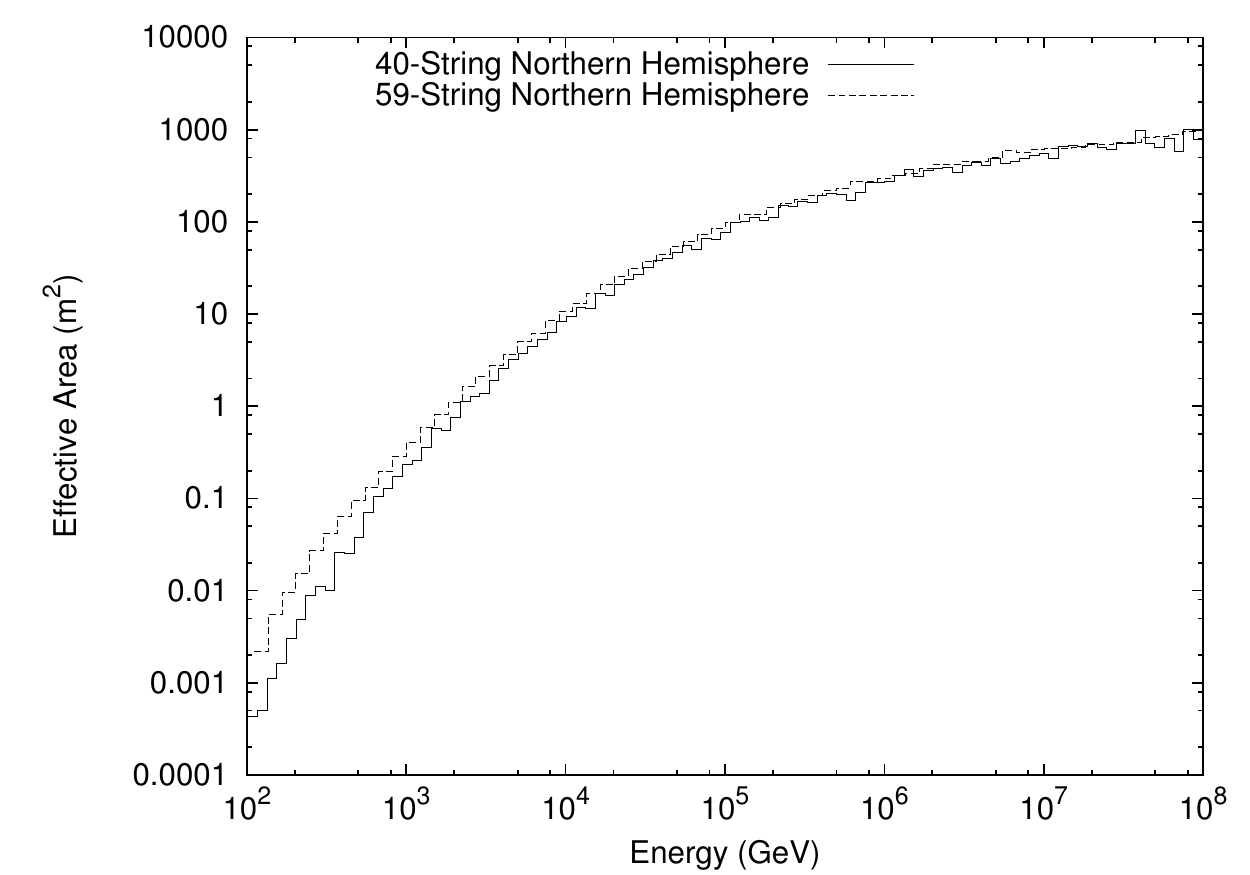}
\caption{Comparison of IceCube effective area between the 40- and 59-string detectors in the northern hemisphere using the model-independent event selection. The increase in effective area is related to the increased geometric area of the detector, as well as to improved event selection and reconstruction techniques. Low-energy sensitivity was especially enhanced as a result of a more favorable detector geometry and the deployment of the first of the strings of the Deep Core subdetector. As a result, the integrated effective area of the 59-string detector for this analysis is more than 1.5 times that of the 40-string detector.}
\label{fig:effarea_ic40ic59}
\end{figure}

\section*{Combination of Datasets}

The results presented use a combination of the IceCube 40- and 59-string datasets. In both analyses, all GRBs were individually simulated and this simulation was applied to the detector running at the time of the GRB. The simulated events from the full GRB catalog were treated as a combined dataset, which was then compared to the combined result from both detector configurations. Systematic uncertainty estimates computed by variation of detector parameters were likewise applied separately to each detector configuration and GRB sub-catalog, and then used in a combined fit for the determination of the final limits.
\clearpage

\end{document}